\documentclass[preprint,showpacs,amsmath,amssymb]{revtex4-2}

\usepackage{amsmath}
\usepackage{graphicx}
\usepackage{amsfonts}
\usepackage{dcolumn}
\usepackage{bm}
\usepackage{color}

\begin{document}

\title{Uncovering the underlying mechanisms of phase transitions in chiral active particles}

\author{Dongrun Jian$^{1,2}$}
\author{Jie Su$^{2}$}
\thanks{E-mail: sj0410@ucas.ac.cn}
\author{Jun Wang$^{1}$}
\thanks{E-mail: wangj@nju.edu.cn}
\author{Jin Wang$^{2,3}$}
\thanks{E-mail: jin.wang.1@stonybrook.edu}
\affiliation{1.Department of Physics, Nanjing University, Nanjing, Jiangsu 210093, China\\2.Center for Theoretical Interdisciplinary Sciences, Wenzhou Institute, University of Chinese Academy of Sciences, Wenzhou 325001, China\\3.Department of Chemistry and of Physics and Astronomy, State University of New York of Stony Brook, Stony Brook, New York 11794, USA}
\date{\today}

\begin{abstract}

Chiral active matter widely exists in nature and emerges rich dynamical behaviors. Among these, chiral active particles (CAPs) with alignment effects show novel collective motions such as orderly rotating droplets and distinct phase transitions under different chirality degrees. However, the underlying dynamical and thermodynamical mechanisms of the phase transitions in the CAP system are not quite clear. Here, by combining the nonequilibrium physics with the coarse-grained mapping method, we quantified the potential landscape and the flux field to reflect global driving forces of the CAP system, characterizing the number and location of the steady states.
Moreover, we revealed that mean flux and entropy production rate are
respectively the dynamical and thermodynamical origins for the nonequilibrium phase transition, further providing a practical tool to confirm the continuity of the phase transition and the phase boundary. Our findings may inspire the design of experimental CAPs and present a new approach for investigating phase transition behaviors in other complex active systems.

\end{abstract}

\maketitle
\section{INTRODUCTION}

Active matter has the ability to take in and convert energy into directed motion, leading to the system far away from equilibrium\cite{re1} and breaking the detailed balance (DB) at the microscopic scale\cite{re2.1,re2.2,re2.3}.
There are a large number of active matter existing in various scales in nature, ranging from nanoscale and microscale, such as molecular motors, sperms, and E. coli, to macroscale, such as fish, birds, and herd\cite{re4.1,re4.2,re4.3,re4.4}. Due to the introduction of the nonequilibrium, active matter system shows many novel collective behaviors, including motility-induced phase separation\cite{re5.1,re5.2,re5.3,re6}, clustering\cite{re6}, turbulence\cite{re7.1,re7.2}, vortex\cite{re8}, self-assembly behavior\cite{re9}, swarm\cite{swarm1,swarm2}, etc. Among various kinds of nonequilibrium physical theories and simulated models \cite{re3.1,re3.2,re3.3}, the Vicsek model (VM) with alignment effects \cite{re4.4,re10} is a significant one to study those behaviors mentioned above, especially swarm appeared in birds and herd \cite{re4.4}. 
To understand more complex behaviors, VM is revised and expanded to numerous different forms, such as 3D Vicsek particles (VPs) \cite{vm1}, VPs moving in the obstacle environment\cite{vm2}, VPs with vectorial noise\cite{vm3}, etc.

Actually, many active particles move chirally such that their motion trajectories are circular. They are so-called dynamically chiral particles, which widely exist in nature, such as biological swimmers (spermatozoa, E. coli, etc.) \cite{cap1, cap3, cap2, cap4, cap5, cap6} or manmade objects (asymmetrical Janus particles) \cite{Janus1}.  Liebchen B. et al introduced dynamically chirality into VM, and the generalized model is called the ``chiral active particle model (CAP model)". The CAP system can spontaneously emerges macro-droplets and micro-droplets patterns, which are totally different from those flocking and band phases that appeared in the original VM system. When the frequency of particles in the CAP model does not hold constant, but rather follows a distribution, the whole active system can emerge vortex phase\cite{CAPm2}, cloud-like structures\cite{CAPm3}, and self-sorting\cite{CAPm4.2,CAPm4.3}. 
The CAP system is one of the most representative models abstracted from various kinds of interesting collective behaviors, rich phases and phase transitions of chiral active matter. However, the underlying dynamical and thermodynamical mechanisms of the phase transition in CAP system, as well as the global physical picture, are still not quite clear and require understanding.

In this study, we first reveal distinct phase transitions dependent on the noise intensity $\eta$ at different particles angular velocity $\omega$ of the CAP. This is the continuous phase transition from the macro-droplet phase, followed by the micro-droplet one, and finally to the uniform disordered one at $\omega=0.5$, while the discontinuous transition can emerge from the micro-droplet phase to the uniform disordered one at $\omega=3.0$. Then, by using the coarse-grained mapping method \cite{map1} combined with the nonequilibrium physics, which is widely used in the nonequilibrium complex and biological systems \cite{LFneural,pf1,LFcellc,LFbio,LFevo,LFeco}, we can quantify the landscape along with flux fields in the phase space involving the local density and local alignment degree of the CAP system.
The potential landscape  not only reflects the global information of the system including the number and weight of the steady states, but also verifies the continuity of the order-disorder phase transition by the change of landscape shapes. It is found that the nonequilibrium flux divides into two groups with opposite directions in the ordered phases with two basins of attraction, which tends to merge these two basins into one and offers the dynamical driving force. Moreover, we quantify the mean flux and the entropy production rate. We find that they both have sharp/gentle variations at the transition threshold of noise intensity $\eta$ under different particle angular velocity $\omega$. This not only reflects the dynamical and thermodynamical origins of the phase transition, but also demonstrates the continuity of transition. Finally, both the nonzero average difference of cross correlation functions and the non-overlapping transition paths in the droplet phases, quantify the degree of the time-reversal symmetry breaking in the ordered phases.

\section{Model and Method}
For $N$ pointlike chiral active particles (CAP) in a 2D box with size $L$ and periodic boundary condition, the motion of the \emph{i-th} CAP satisfies the following Langevin equations,
\begin{equation}\begin{aligned}
		\dot{\mathbf{x}}_i &=\nu\mathbf{n}_i \\
		\dot{\theta_{i}} &=\omega+\frac K{\pi R_\theta^2}\sum_{j\in\partial_i}\sin(\theta_j-\theta_i) +\xi_i 
\end{aligned}\end{equation}

\noindent Herein, $x_{i}, y_{i}$, and $\theta_{i}$ are respectively the horizontal coordinate, vertical coordinate, and angle of the \emph{i-th} particle. 
$\mathbf{x}_i$, $\nu$, and $\mathbf{n}_i=(\cos\theta_i,\sin\theta_i)$ are respectively the position, the self-propulsion velocity and the moving direction of the \emph{i-th} particle. $\omega$ is the angular velocity of the particle, representing the chirality of CAP. 
The second term in the rotation equation represents the alignment interaction of the \emph{i-th} particle with $K$ the interaction strength and $R_{\theta}$ the interaction radius. 
Noise term $\xi$ is independent Gaussian white noise with zero mean and time correlation $\langle\xi_i(t)\xi_j(t^{\prime})\rangle=2\eta\delta_{ij}\delta(t-t^{\prime})$, where $\eta$ is the rotational diffusion coefficient. 

In our simulation, we fix the number of particles $N$=32000, the size $L$=80, the interaction radius $R_\theta$=1, and the time step $\Delta t$=0.1. The initial construction of CAP system is random. 
We first simulate for $1\times 10^6$ time steps to ensure the system reaches the  nonequilibrium steady state and we perform the simulation in another $1\times 10^6$ time steps in the steady state for sampling.

To further explore the dynamical and thermodynamical mechanisms of the phase transition, we combine the nonequilibrium physics \cite{pf1} with the coarse-grained mapping \cite{map1} for exploration. 
First, we divide the real space into M$\times$M cells of the same size, whose information can be characterized by the local density $\rho_i$ and the local alignment degree $\phi_i=|\sum_{q\in cell_i}\mathbf{v}_q|/(n_i\nu)$, where $\mathbf{v}_q$ is the speed of \emph{q-th} CAP in the \emph{i-th} cell, and $n_i$ is the number of CAPs in the \emph{i-th} cell. Then the information in the real space can be mapped into the $\rho-\phi$ phase space. Based on this mapping method, we can calculate the probability distribution $P(\mathbf{x},t)$ with $\mathbf{x}=(\rho,\phi)^T$. (More details can be found in the supplemental information, SI).
Moreover, the Langevin equations in $\rho-\phi$ phase space can be written as: 
\begin{equation}\begin{aligned}
		\dot{\rho}(\mathbf{x},t) &=F_{\rho}(\mathbf{x})+\zeta_{\rho}(\mathbf{x},t) \\
		\dot{\phi}(\mathbf{x},t) &=F_{\phi}(\mathbf{x})+\zeta_{\phi}(\mathbf{x},t)\\
\end{aligned}\end{equation}
Herein, $\mathbf{F}=(F_{\rho}, F_{\phi})^T$  is the deterministic ``driving forces'' and $\mathbf{\zeta}=(\zeta_{\rho}, \zeta_{\phi})^T$ is the stochastic force with time correlation $\langle\mathbf{\zeta}(\mathbf{x},t)\mathbf{\zeta}(\mathbf{x},t^\prime)\rangle=2\mathbf{D}(\mathbf{x})\delta(t-t^\prime)$, where $\mathbf{D}(\mathbf{x})$ is a $2\times 2$ diffusion coefficient tensor. Both $\mathbf{F}$ and $\mathbf{D}$ can be obtained by the coarse-grained mapping method (More details can be found in the SI).

Then we can apply the nonequilibrium potential and flux theory \cite{pf1} (details can be found in the SI) in the $\rho-\phi$ phase space. When the CAP system reaches the steady state, $P(\mathbf{x},t)$ is independent of time $t$, i.e., $\partial P_{ss}(\mathbf{x})/\partial t=0$, where the subscript ss denotes the steady state. We can define the effective nonequilibrium potential landscape $U_{neq}$ naturally as: 
\begin{equation}
		U_{neq}(\mathbf{x}) =-\ln(P_{ss}(\mathbf{x}))
\end{equation}
On the other hand, the steady state nonequilibrium flux $\mathbf{J}_{ss}$ is satisfied with $\nabla\cdot\mathbf{J}_{ss}=0$, due to the Fokker-Planck equation $\partial P/\partial t=-\nabla\cdot\mathbf{J}$ (More details can be found in SI). Therefore, $\mathbf{J}_{ss}$ can be written as:
\begin{equation}
	\mathbf{J}_{ss}(\mathbf{x})=\mathbf{F}(\mathbf{x})P_{ss}(\mathbf{x})-\nabla\cdot(\mathbf DP_{ss}(\mathbf{x})).
\end{equation}
The driving force $\mathbf{F}$ for nonequilibrium systems are determined both by the gradient of the potential landscape $U_{neq}$ and the rotational steady state probability flux $\mathbf{J}_{ss}$ (More details can be found in SI), $\mathbf{F}=\mathbf{F}_{grad}+\mathbf{F}_{curl}=-\mathbf {D}\nabla U_{neq}+{J}_{ss}/P_{ss}$. For the simplicity of notations in the following, we remove the corner notation from $\mathbf{J}_{ss}$, $P_{ss}$, and $U_{neq}$.

\section{Result and discussion}
\subsection*{Dynamical phase diagram of the CAP system}
\begin{figure}[h]
\begin{centering}
\centering
\includegraphics[width=1.0\columnwidth]{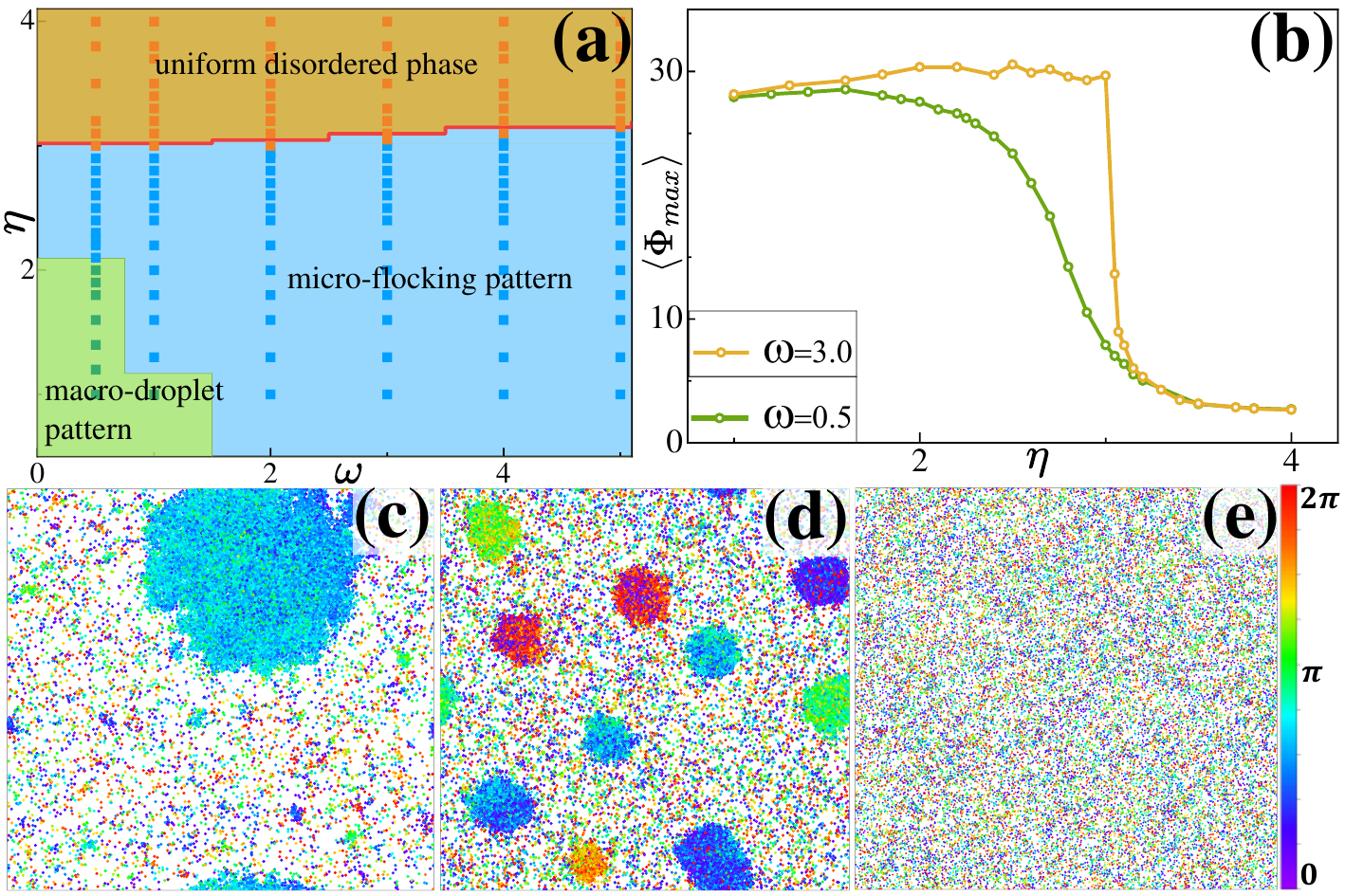}
\par\end{centering}
\caption{Phase diagram, typical snapshots, and dependence of $\Phi_{max}$ on $\eta$. (a) The phase diagram of the CAP system in the $\omega-\eta$ plane. The red line is the phase boundary between the disordered phase (orange area) and the ordered phases (the macro-droplet phase, green area, and the micro-droplet phase, blue area). (b) The variation of $\Phi_{max}$ with $\eta$. Simulation snapshots of the macro-droplet pattern (c) with $\omega=1.0$, $\eta=1.0$, the micro-droplet pattern (d) with $\omega=5.0$, $\eta=2.0$, and disordered phase (e) with $\omega=3.0$, $\eta=4.0$. The colors of CAPs denote their orientations.  
}
\label{pd}
\end{figure}

First of all, we perform simulations under different parameters to generate the steady states of the CAP system. Interestingly, the CAP system can spontaneously form macro-droplets and micro-droplets. Considering the significance of noise intensity in the Vicsek model and the importance of chirality degree in CAPs, we fix $\nu=0.55$, $K=0.8$, and then obtain a phase diagram based on $\eta$ and $\omega$ as shown in Fig.~\ref{pd}(a).

Three distinct dynamical phases can be found in this phase diagram: Specifically, the uniform disordered phase (Fig.~\ref{pd}(e)) occupies the region characterized by large $\eta$, whereas the two ordered phases are situated in the area of small $\eta$. As an example shown in Fig.~\ref{pd}(c), the macro-droplet phase consists of a single big, rotating cluster with CAPs moving in similar directions, which is located in the region with small $\omega$ and $\eta$. Conversely, the micro-droplet phase comprising numerous small, orderly rotating droplets (Fig.~\ref{pd}(d)) is situated at relatively large $\eta$ and $\omega$. These findings suggest that CAPs tend to spontaneously organize into a big, orderly rotating cluster due to their alignment effect and dynamical chirality. However, increased noise intensity and faster rotation of CAPs lead to the decomposition of this large cluster into smaller droplets, ultimately resulting in the transition to the disordered phase.

Since the macro-droplet phase only occurs at small $\omega$, we then focus on the two kinds of phase transitions dependent on $\eta$ at $\omega=0.5$ (with the macro-droplet phase)  and $\omega=3.0$ (without the macro-droplet phase), respectively. To describe these transitions more clearly, we introduce an ordered parameter $\Phi_{max}=max\{\rho_i\phi_i\}$.
The obtained $\Phi_{max}$ vs $\eta$ is shown in Fig.~\ref{pd}(b), we can ensure that the phase transition is discontinuous (``first order") at $\omega=3.0$ (yellow line), while it becomes continuous (``second order") in the case of $\omega=0.5$ (green line).

\subsection*{Global information based on the potential landscape}

\begin{figure}[h]
	\begin{centering}
		\centering
		\includegraphics[width=1.0\columnwidth]{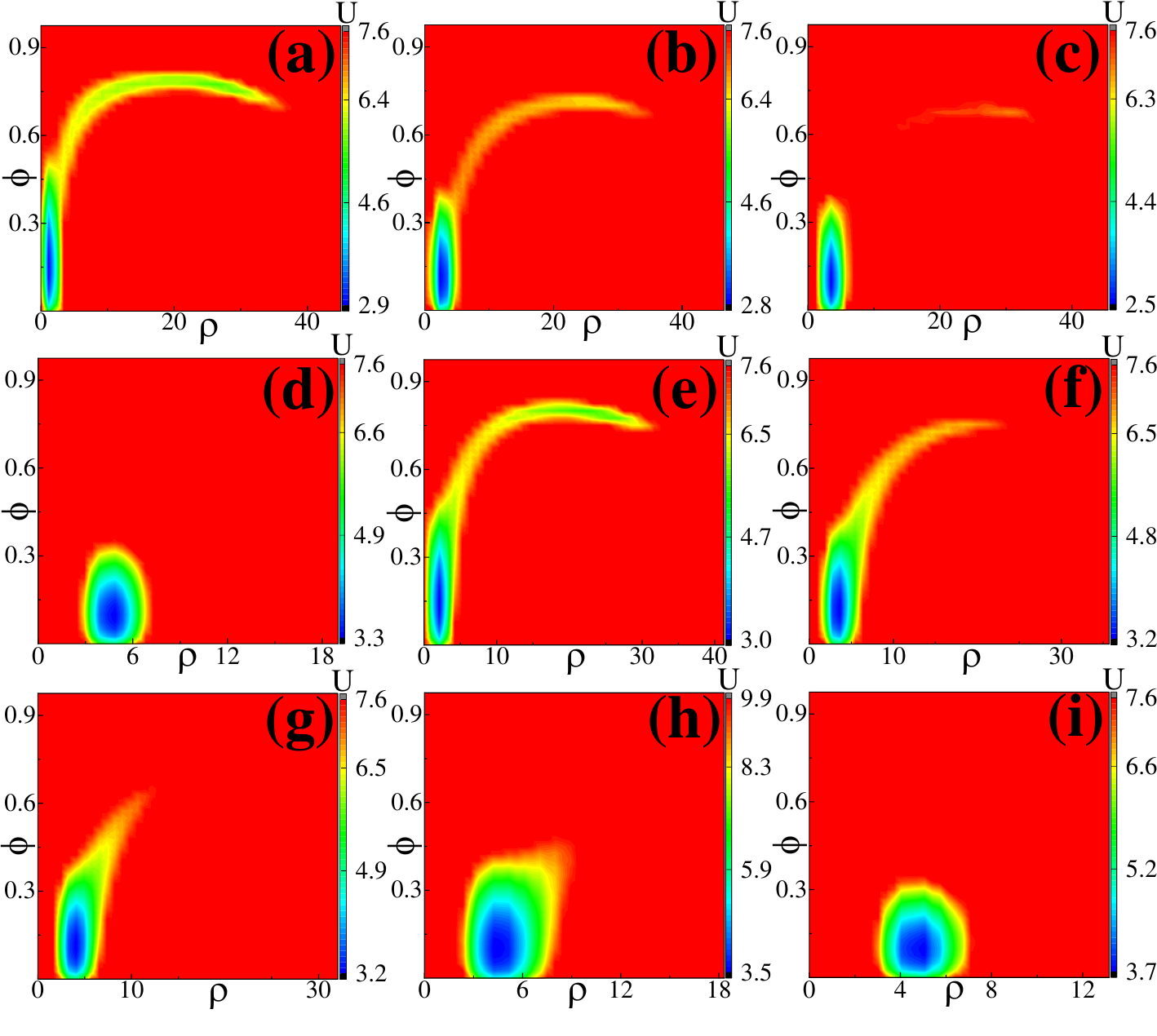}
		\par\end{centering}
	\caption{Typical nonequilibrium potential (colored background) of the CAP system in phase space with different $\eta$ and $\omega$. (a)-(d) : $\omega$=3.0, $\eta$=1.8(a), $\eta$=2.6(b), $\eta$=2.9(c), $\eta$=3.1(d). (e)-(i) : $\omega$=0.5, $\eta$=1.8(e), $\eta$=2.4(f), $\eta$=2.6(g), $\eta$=2.9(h), $\eta$=3.1(i).}
	\label{pt}
\end{figure}

By combining the coarse-grained mapping method and the nonequilibrium potential and flux theory, we can first quantify the global information of the CAP system via the potential landscape as shown in Fig.~\ref{pt}.

For small $\eta$ at both $\omega$, we can see two potential wells with one located at large $\rho$ and $\phi$ (representing the ordered flocking state) and another one situated at small $\rho$ and $\phi$ (denoting the disordered dilute state). In the case of $\omega=3.0$ (Fig.~\ref{pt}(a)-(d)), it is observed that the potential basin is located at large $\rho$ and $\phi$ disappears suddenly at $\eta_{pt}\equiv\eta=3.0$, indicating that the system undergoes a discontinuous phase transition.

However, for the situation of $\omega=0.5$ (Fig. ~\ref{pt}(e)-(i)), it is found that the potential basin located at large $\rho$ and $\phi$ first begins to flatten out, then becomes a long-tail extension of the other basin, and ultimately evolves into one single basin. This smooth alteration in the potential landscape indicates a continuous phase transition occurring at $\omega=0.5$. 
These results about the transition continuity show the same conclusion as that in the aforementioned Fig~\ref{pd}(b).

\subsection*{The nonequilibrium flux inside the landscape}

In addition to the potential landscape of the CAP system, the nonequilibrium flux field induced by the driving force can be obtained from the nonequilibrium potential and flux theory\cite{LFbio}. When the system forms the uniform disordered phase at large $\eta$ for both large $\omega$ and small $\omega$, it is observed that the flux mainly flows towards the single potential basin (see Fig. S3 in the SI). Under this situation when noise is large enough, the changes of $\rho$ and $\phi$ are contributed by stochastic fluctuations on the whole. 

\begin{figure}
\begin{centering}
\centering
\includegraphics[width=1.0\columnwidth]{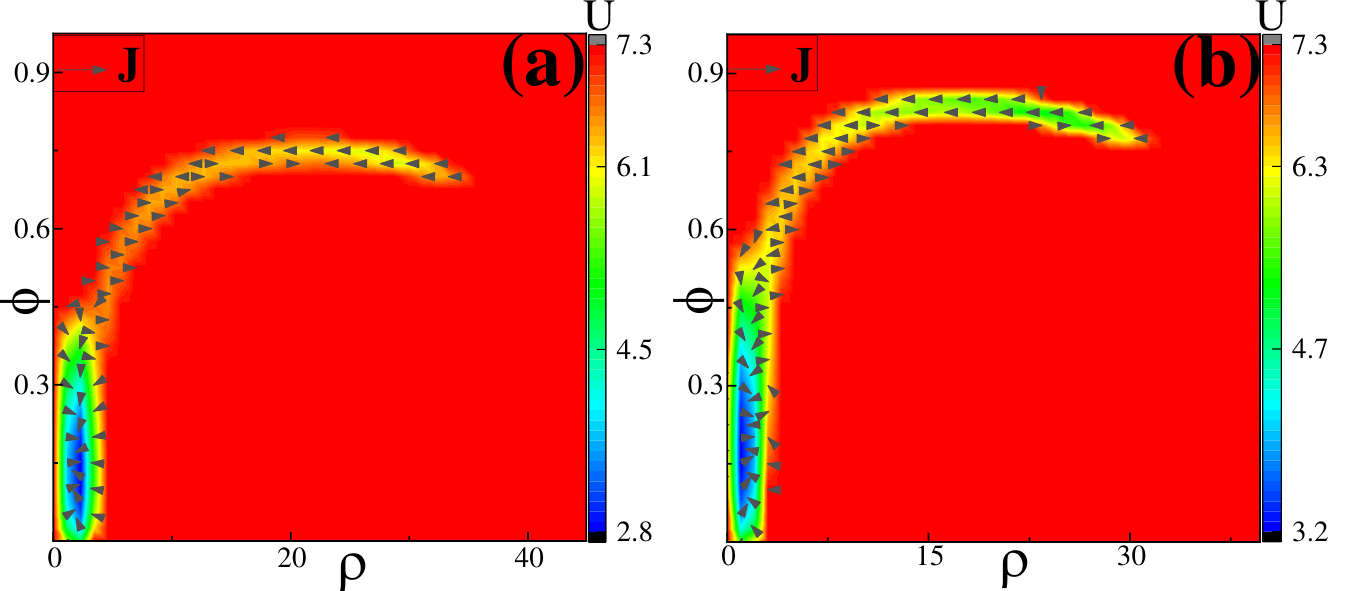}
\par\end{centering}
\caption{Typical nonequilibrium potential (colored background) and flux $\mathbf{J}$ (gray arrows) of the CAP system in phase space. (a): $\omega$=3.0, $\eta$=2.0. (b):$\omega$=0.5, $\eta$=1.4.}
\label{uj}
\end{figure}

However, in the case of the micro-droplet pattern with $\omega = 3.0$, $\eta = 2.0$ (Fig.~\ref{uj}(a)), it is found that there are two groups of fluxes connecting the two potential basins with one (named as high density basin) located at the high $\rho$-$\phi$ area, representing the flocking state consisting of the small rotating droplets; and the other one (named as low density basin) situated at the low $\rho$-$\phi$ region, indicating the disordered gas state. Remarkably, these two groups of fluxes point to opposite directions, i.e., one group of fluxes streams from low density basin to high density basin, while the other group of fluxes is in reverse. For the situation of the macro-droplet pattern with $\omega = 0.5$, $\eta = 1.4$ (Fig.~\ref{uj}(b)), it can be observed that the fluxes have similar behaviors. More interestingly, it is found that for both micro-droplet and macro-droplet patterns, the fluxes inside the high density basin mainly point to the negative $\rho$-direction, meaning that the nonequilibrium flux provides the dynamical driving force to merge these two potential basins into one, further leading to the phase transition from the ordered phases to the disordered phase.

Different from the gradient force always pointing to the potential basin, these nonequilibrium fluxes divide into two groups with opposite directions, in favor of the switching between the low and high density basins. In other words, the nonequilibrium flux prefers global movements rather than attracting into the potential basin. This not only provides the dynamical driving force of the phase transition from two potential basins to one basin, but also becomes the main element that drives the system far away from equilibrium state breaking the detailed balance. To characterize this detailed balance breaking, we quantify the transition paths between these two potential basins (see Fig. S4 in the SI). It is found that these pairs of transition paths do not overlap with each other, reflecting the breaking of time-reversal symmetry.

\subsection*{The dynamical and thermodynamical mechanisms of the phase transition}

Now, we try to describe the dynamics and thermodynamics of the CAP system and understand the underlying mechanisms of the phase transition in the CAP system. Here, we respectively introduce the mean flux and the entropy production rate (EPR) to measure the dynamical and thermodynamical natures of the system.

\begin{figure}
	\begin{centering}
		\centering
		\includegraphics[width=1.0\columnwidth]{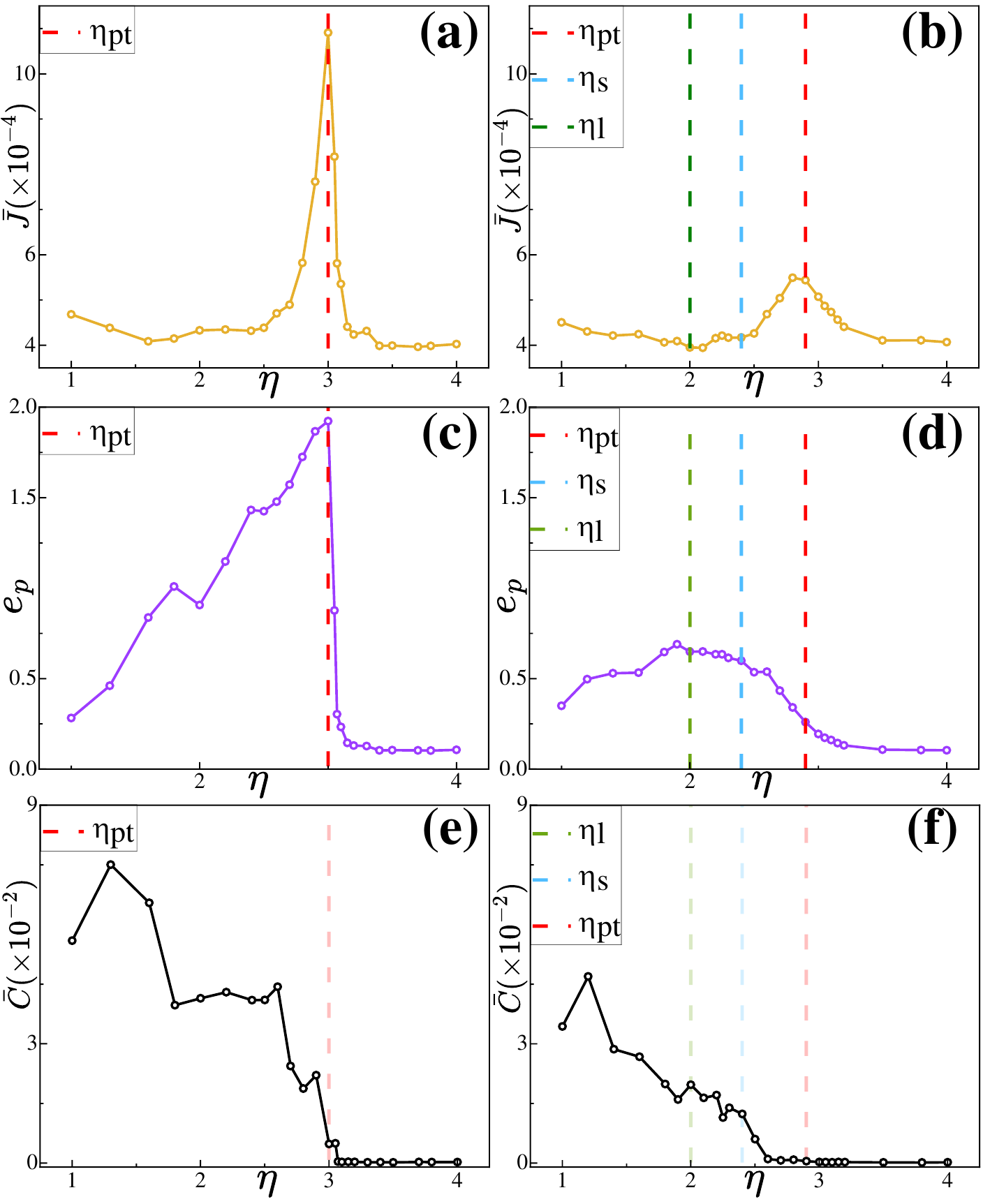}
		\par\end{centering}
	\caption{Dependence of mean flux $\bar{J}$ [(a), (b)], the entropy production rate $e_p$ [(c), (d)] and the average difference of cross correlation functions $\bar{C}$ [(e), (f)] on $\eta$. The red line $\eta_{pt}$ is the boundary between disordered phase and flock phase, and the dashed green line $\eta_l$ and blue line $\eta_s$ denote the threshold between macrodroplet and microflock pattern. (a), (c), (e) are in fast rotating speed, $\omega=3.0$. (b), (d), (f) are in slow rotating speed, $\omega=0.5$.}
	\label{cur}
\end{figure}

The contribution of the nonequilibrium flux can be  measured by mean flux $\bar{J}$, which is defined as: 
\begin{equation}
	\bar{J}\equiv\int|\mathbf{J}|\mathrm{d}\mathbf{x}/(\int\mathrm{d}\mathbf{x}).
\end{equation}

The obtained $\bar{J}$ dependent on the noise intensity $\eta$ for both large and small $\omega$ are plotted in Fig.~\ref{cur}(a) and (b), respectively. Remarkably, it is observed that for both situations,  $\bar{J}$ shows a peak at the boundary ($\eta_{pt}$) between the ordered phase and the disordered phase. The steady state flux being curl tends to rotate around instead of being localized at current state. This can destabilize the current state. It can lead to the bifurcation/phase transition. Thus, flux can serve as the dynamical origin of critical transitions. The appearance of these peaks not only demonstrates that the nonequilibrium flux as being the dynamical origin of the order-disorder phase transition, but also provides an appropriate tool to confirm the phase boundary. More interestingly, Compared to the case for $\omega=3.0$ (Fig.~\ref{cur}(a)), the peak of $\bar{J}$ in the one for $\omega=0.5$ (Fig.~\ref{cur}(b)) is wider and shorter, implying that the order-disorder phase transition in the CAP system with $\omega=3.0$ is sharper, i.e., the transition for $\omega=3.0$ is more violent than that for $\omega=0.5$. Moreover, for the situation with $\omega=0.5$ (Fig.~\ref{cur}(b)), $\bar{J}$ exhibits different behaviors at different phases: as $\eta$ increases, it decreases in the macro-droplet phase before the threshold $\eta_l$ but increases in the micro-droplet phase beyond $\eta_l$. More interestingly, the increasing rate of $\bar{J}$ in the micro-droplet phase shows different slopes separated by the blue dashed line ($\eta_s$). Via detailed analysis, it is found that the small droplets are in low $\phi$ when $\eta>\eta_s$, but a relatively large droplet in high $\phi$ spontaneously forms when $\eta\leq\eta_s$ (see typical snapshots of Fig. S5 in the SI). In other words, as $\eta$ increases, the CAP system at $\omega=0.5$ will undergo macro-droplet, ordered micro-droplet, disordered micro-droplet, and uniform disordered patterns in turns.

Then we focus on the view of thermodynamics and quantify the EPR as follows: 
\begin{equation}
e_p\equiv\int P^{-1} \mathbf{J}^\mathrm{T} \mathbf{D}^{-1} \mathbf{J} \mathrm{d}\mathbf{x}.
\end{equation}

For $\omega$=3.0[Fig.~\ref{cur}(c)], $e_p$ first increases gradually with increasing $\eta$, then reaches its maximum at $\eta_{pt}$ and is followed by a sharp decline.
For $\omega=0.5$ [Fig.~\ref{cur}(d)], the behavior of $e_p$ dependent on $\eta$ becomes more complicated. First, the peak of $e_p$ shifts to $\eta_l$, and the decline after the peak turns to be more gradual, demonstrating again that the order-disorder phase transition in the CAP system with $\omega=3.0$ is more violent. Additionally, the decrease of $e_p$ also shows different slopes in the ordered micro-droplet, disordered micro-droplet, and uniform disordered patterns. Entropy production rate is closed linked to the flux. It represents the thermodynamic dissipation cost. Since the flux can serve as the dynamical origin of critical transitions, the EPR can provide the thermodynamic origin of the critical transitions. Therefore, we argue that the EPR can serve as the thermodynamical origin of the phase transition emerging in the CAP system.

Finally, we introduce the average difference ($\bar{C}$) between the two-point cross correlations forward ($C_1$) and backward ($C_2$) in time to measure the time irreversibility of the system, which is defined as $\bar{C}\equiv \sqrt{\int^{T}_0(\langle C_1(\tau)-C_2(\tau)\rangle_x)^2\mathrm{d}\tau/T}$, where $C_1(\tau)=\langle\rho(t)\phi(t+\tau)\rangle_t$, $C_2(\tau)=\langle\rho(t+\tau)\phi(t)\rangle_t$, and $\langle A \rangle_x$ means the ensemble average of $A$ over cells. 
The obtained $\bar{C}$ dependent on $\eta$ for the CAP system with $\omega=3.0$ and $\omega=0.5$ are plotted in Fig.~\ref{cur}(e) and (f), respectively. It is found that for both cases, $\bar{C}$ decreases with the increasing $\eta$ before the threshold $\eta_{pt}$, and maintains a constant close to $0$. These results reflect the breaking of time-reversal symmetry in the macro-droplet and micro-droplet phases. Moreover, the change behavior of $\bar{C}$ at $\eta_{pt}$ can provide early warning signals of the order-disorder phase transition.

\section{CONCLUSION}

In summary, we establish the potential and flux landscape of the CAP system in the $\rho$-$\phi$ phase space by using the coarse-grained mapping method combined with the nonequilibrium physics. 
Both $\bar{J}$ and $e_p$ show variations at the transition threshold, not only demonstrating that $\bar{J}$ and $e_p$ are respectively the dynamical and thermodynamical origins of the phase transition, but also providing a practical tool to confirm the phase boundary.
More interestingly, both the variation intensity of $\bar{J}$ and $e_p$ at the transition threshold, as well as the shape change of the potential landscape, can be used to estimate the continuity of the phase transition.
Furthermore, time-reversal symmetry breaks down in the droplet phase, reflected by non-overlapping transition paths and nonzero $\bar{C}$.
Our findings not only reveal the dynamical and thermodynamical nature of the CAP system, which may provide potential applications to design experimental CAPs, but also offer a new way to explore phase transition behaviors in other complex active systems.

\section*{ACKNOWLEDGMENTS}
R.J. and J.S. are supported by WIUCASQD2022012, NSFC12404237 and NSFC12234019.
\section*{DATA AVAILABILITY}
All study data are included in this article and/or SI Appendix.
\section*{COMPETING INTERESTS}
The authors declare no competing interests.
\section*{AUTHOR CONTRIBUTIONS STATEMENT}
Jin Wang and Jun Wang conceived the idea and designed the research. Jie Su provided the model and supervised the research. Dongrun Jian performed the simulations and data analysis. All authors discussed the results and co-wrote the manuscript.


\end{document}